\documentclass[twocolumn,pra,aps,superscriptaddress]{revtex4-1}
\usepackage[english]{babel}
\usepackage[utf8]{inputenc}
\usepackage{url,psfrag,graphicx}
\usepackage{dcolumn}
\usepackage{amsmath,amssymb}
\usepackage{bm}
\usepackage{hyperref}
\usepackage{epsfig}

\def\<{\left<}
\def\>{\right>}
\def\ket|#1>{\left|#1\right>}
\def\bra<#1|{\left<#1\right|}
\def\elem<#1|#2|#3>{\left<#1\right|#2\left|#3\right>}
\def\braket<#1|#2>{\langle#1|#2\rangle}
\def\({\left(}
\def\){\right)}
\def\[{\left[}
\def\]{\right]}

\def\Tr{\hbox{Tr}}

\def\beq{\begin{equation}}
\def\eeq{\end{equation}}
\def\mat#1,#2,#3,#4.{\begin{pmatrix}#1&#2\\#3&#4\end{pmatrix}}
\def\matt#1,#2,#3,#4,#5,#6,#7,#8,#9.{\begin{pmatrix}#1&#2&#3\\#4&#5&#6\\#7&#8&#9\end{pmatrix}}
\def\fg{\textrm{fg}}
\def\ag{\textrm{ag}}
\def\fb{\textrm{fb}}
\def\ab{\textrm{ab}}
\newcommand\undermat[2]{% http://tex.stackexchange.com/a/102468/5764
  \makebox[0pt][l]{$\smash{\underbrace{\phantom{%
    \begin{matrix}#2\end{matrix}}}_{\text{$#1$}}}$}#2}

\begin{document}

\title[Short Title]{Efficient computation of matrix elements of
  generic Slater determinants} 

\author{Javier Rodriguez-Laguna}
\affiliation{Dpto. F\'{\i}sica Fundamental, Universidad Nacional de
  Educaci\'on a Distancia (UNED), Madrid, Spain.}

\author{Luis Miguel Robledo}
\affiliation{Dpto. F\'{\i}sica Te\'orica, Universidad Aut\'onoma de
  Madrid (UAM), Madrid, Spain.}
\affiliation{Center for Computational Simulation, Universidad Polit\'ecnica de Madrid,
  Campus de Montegancedo, 
  Boadilla del Monte, 28660-Madrid, Spain}

\author{Jorge Dukelsky}
\affiliation{Instituto de Estructura de la Materia (IEM), CSIC,
  Madrid, Spain.}

\date{September 9, 2019}

\begin{abstract}
We present an extension of the L\"owdin strategy to find arbitrary
matrix elements of generic Slater determinants. 
The new method applies to arbitrary number of fermionic operators, even in the case of a singular overlap matrix. 
\end{abstract}

\maketitle

%%%%%%%%%%%%%%%%%%%%%%%%%%%%%%%%%%%%%%%%%%%%%%%%%%%%%%%%%%%%%%%%%%%%%%%%%

\section{Introduction}
Many developments in quantum many body physics require the efficient
computation of matrix elements of fermionic operators between Slater
determinant states.  In many relevant cases, the matrix elements
should be computed between Slater determinant states which are not
based on the same set of single-body fermionic states. An early
example of this kind of calculations can be found in the seminal paper
by the Swedish physicist Per-Olov L\"owdin \cite{PhysRev.97.1490} who developed in 1955 a
very smart strategy, based on a careful application of the properties
of the determinants. The strategy was developed in full for the case
of two and four fermionic operators, and required the overlap matrix
between the two Slater determinants to be non-singular
\cite{PhysRev.97.1490}.  Further developments, aimed to simplify the
original complexity of the formulas and to facilitate their use in the
framework of the valence bond theory, can be found in the literature
\cite{Prosser68,PhysRevA.31.2107}. There are many applications in
quantum chemistry requiring overlaps of operators between Slater
determinant including Configuration Interaction (CI) and symmetry
restoration methods (see Refs
\cite{KOCH1993193,VERBEEK1991115,TOMITA1996687,Scuseria11,Jimenez-Hoyos2013,Rodriguez-Guzman2014}
as an example).  More recently Brouder \cite{PhysRevA.72.032720}
proposed a method to reduce the combinatorial complexity of Wick's
theorem to a more manageable algebraic complexity. As applications of
the method, formulas for the overlap of a general product of creation
and annihilation operators between arbitrary Slater determinants were
proposed and used to compute, for instance, the generating function of
the Green function or $k$-density correlation operators. The method
uses general ideas coming from the world of quantum groups and Hopf
algebras but leads to rather involved expressions. Slater determinants
are also used to expand the wave functions of the fractional quantum
Hall effect (FQHE) as discussed in Ref \cite{PhysRevLett.103.206801}.
In lattice QCD the study of physical systems involving several hadrons
\cite{Detmold2013} as it is the case in the description of collisions
\cite{PhysRevD.81.111504} or when the hadrons aggregate to form atomic
nuclei \cite{Beane2011,Savage2012,Doi2013117}, require the evaluation
of matrix elements involving the product of $(3N)^{2}$ ($N$ is the
number of hadrons) creation or annihilation operators coming from the
quark fields. If the matrix elements are computed in terms of Wick
contractions (see below), the required number of terms grows
exponentially fast with $N$ \cite{Detmold2013}. General overlaps of
Slater determinants are also required in nuclear physics, in the
framework of the MonteCarlo Shell Model (MCSM)
\cite{OTSUKA2001319,UTSUNO2013102}, or in the field of symmetry
restored quasiparticle excitations \cite{SCHMID2004565,Puddu06}.

In all the above situations the matrix elements can be computed with the
help of Wick's theorem or its generalizations, but the number of contractions
to consider grows with the factorial of the number of operators involved and 
therefore it becomes unmanageable very soon.  

In this work we extend L\"owdin's results to the case of a generic
number of fermionic operators in order to obtain compact and easily 
handled expressions prone to an efficient evaluation in a computer. In
addition, the evaluation of hundred of thousands, if not millions of
operator overlaps, calls for robust evaluation methods capable to
handle the cases of zero or nearly zero overlaps of the Slater
determinants where L\"owdin's method becomes ill-defined. In our derivation we will use
a second quantization formalism from the beginning, which makes
calculations more transparent.

This article is organized as follows. Our generalized version of
Löwdin's theorem is described and proved in Sec. \ref{sec:gwt}. The
case of zero overlap between both Slater determinants is discussed in
Sec. \ref{sec:zero}. A numerical application is provided in
Sec. \ref{sec:numerical}, where our approach is used to estimate the
entanglement of a block for a linear combination of Slater
determinants. The article finishes with some conclusions and our
proposals for further work.

%%%%%%%%%%%%%%%%%%%%%%%%%%%%%%%%%%%%%%%%%%%%%%%%%%%%%%%%%%%%%%%%%%%%%%%%

\section{Generalized Löwdin's Theorem}
\label{sec:gwt}

To start with, let us consider two generic Slater determinants
\begin{align}
  \ket|A> =& a^\dagger_{i_1} \cdots a^\dagger_{i_N} \ket|->, \\
  \ket|B> =& b^\dagger_{j_1} \cdots b^\dagger_{j_N}\ket|->.
  \label{eq:defAB}
\end{align}
of $N$ particles. The $a^\dagger_{i}$ and $b^\dagger_j$ are arbitrary
creation operators with quantum numbers denoted by $i$ and $j$, respectively.
Their hermitian conjugate annihilates the true
Fock vacuum $\ket|->$:
\begin{equation}
  a_i\ket|->=b_j \ket|-> = 0,
  \label{eq:annihilate}
\end{equation}
and such that
\begin{align}
  \{ a^\dagger_i, b_j\} &= S^*_{ij}, \\
  \{ a_i, b^\dagger_j\} &= S_{ij}, 
\end{align}
as well as
\begin{equation}
  \{a^\dagger_i,b^\dagger_j\}=0.
\end{equation}
The overlap matrix is defined by $S_{ij}=\<a_i|b_j\>=\bra<-|a_i
b^\dagger_j\ket|->$. The overlap between both states, $\<A|B\>$ is
evaluated in a recursive way:
\begin{align}
  \<A|B\> = &\bra<-| a_N \cdots a_1 b^\dagger_1 \cdots b^\dagger_N \ket|-> \\
  =& - \bra<-| a_N \cdots a_2 b^\dagger_1 a_1
               b^\dagger_2 \cdots b^\dagger_N \ket|-> \\
  &  + S_{11} \bra<-| a_N \cdots a_2 b^\dagger_2 \cdots b^\dagger_N \ket|->
\end{align}
by {\em jumping} with the $b^\dagger_1$ creation operator over the
$a_1$ annihilation one. The notation has been also simplified by
replacing indexes $i_1, \ldots$ by $1,\ldots$.  Let us now introduce
the quantity
\begin{equation}
  \<A|B\>_{[11]}=\bra<-| a_N \cdots a_2 b^\dagger_2\cdots b^\dagger_N
  \ket|->,
  \end{equation}
that corresponds to the overlap of the two Slater determinants, but
``removing the $a_1b^\dagger_1$ pair from $\<A|B\>$''. Then,
\begin{align}
  \<A|B\> &= S_{11}\<A|B\>_{[11]} -S_{21}\<A|B\>_{[21]} +\cdots \nonumber\\
  & +(-1)^{N+1}S_{N1}\<A|B\>_{[N1]}, 
  \label{eq:minor_expansion}
\end{align}
and the expansion ends after $N$ jumps because $\bra<-|b^\dagger_1 =0$.
We easily recognize in $\<A|B\>_{[11]}$ the minor of $S$ with respect 
to the matrix element $(1,1)$, i.e. $S_{11}$. Viewed from this perspective, 
the expression of $\<A|B\>$ given in Eq. \eqref{eq:minor_expansion}  
becomes the minor expansion of the determinant of $S$ by the first row, i.e.
\begin{equation}
  \<A|B\>=\det(S).
\end{equation}
Let us now expand $a^\dagger_i$ and $b^\dagger_j$ in terms of a common
basis $\{c^\dagger_k,k=1,\ldots,N_B\}$
\begin{align}
  a^\dagger_i &= \sum_{k=1}^{N_B} A_{ki} c^\dagger_k,  \\
  b^\dagger_j &= \sum_{k=1}^{N_B} B_{kj} c^\dagger_k, 
\end{align}
then the $N \times N$ overlap matrix $S$ becomes the product of the
two expansion matrices, $A$ and $B$, of dimension $N_B \times N$
\begin{equation}
  S_{ij}=\sum_k A^*_{ki} B_{kj} = (A^\dagger B)_{ij}.
\end{equation}
The previous result can be easily generalized to the calculation of a general
overlap
\begin{equation}
  \bra<A| f_M \cdots f_1 g^\dagger_1 \cdots g^\dagger_M \ket|B>,
  \label{eq:genover}
\end{equation}
where the $f_{l}$ and $g^{\dagger}_{p}$ are arbitrary annihilation and
creation operators expressed in the ${c^\dagger}$ basis as
\begin{align}
  f_l        &= \sum_{k=1}^{N_B} F_{kl} c_k,  \\
  g^\dagger_p &= \sum_{k=1}^{N_B} G_{kp} c^\dagger_k, 
\end{align}
in terms of the $F$ and $G$ matrices of dimension $N_B \times N$. This kind
of overlaps appear when considering a system of $N+M$ particles where
$M$ of them play a different role than the remaining $N$ ones and
therefore require of a different set of orbitals. 
As the $f$'s anti-conmute with the $a$'s and the $g^{\dagger}$'s with the
$b^{\dagger}$, we can repeat verbatim the previous considerations for
$\<A|B\>$. We only have to be careful and define four partial overlap
matrices:

\begin{align}
  (S_\fg)_{ij} = \bra<-| f_i g^\dagger_j \ket|-> & \qquad (M\times M), \\
  (S_\ag)_{ip} = \bra<-| a_l g^\dagger_j \ket|-> & \qquad (N\times M), \\
  (S_\fb)_{lj} = \bra<-| f_i b^\dagger_p \ket|-> & \qquad (M\times N), \\
  (S_\ab)_{lp} = \bra<-| a_l b^\dagger_p \ket|-> & \qquad (N\times N),
\end{align}
to arrive to the formula
\begin{equation}
  \bra<A| f_M \cdots f_1 g^\dagger_1 \cdots g^\dagger_M \ket|B> =
  \det\mat S_\fg,S_\fb,S_\ag,S_\ab.
\label{eq:gwt_det}
\end{equation}
which is the general result for the overlap of Eq.~(\ref{eq:genover}). 
In order to disentangle the contributions from each set of orbitals it 
is convenient to use the
well known formula for the determinant of a partitioned matrix
\begin{align}
  \det \mat P,Q,R,S. &= \det P \det( S-RP^{-1}Q ) \nonumber\\
  &= \det S \det (P-QS^{-1}R).
\label{eq:det_block}
\end{align}
in order to obtain
\begin{equation}
  \det \mat S_\fg,S_\fb,S_\ag,S_\ab. = \det S_\ab
  \det\(   S_\fg-S_\fb S_\ab^{-1} S_\ag\)
  \label{eq:gwt}
\end{equation}
which we can call {\em generalized Löwdin's theorem} (GLT). It requires the
evaluation of the determinant of one $M\times M$ matrix and the determinant
and inverse of a $N\times N$ matrix. This formula is also advantageous 
over Eq.~(\ref{eq:gwt_det}) when $N\gg M\gg 1$
and many matrix elements for different $g$ or $f$ orbitals are required as
only one costly matrix inversion is required.
% keep
%This compact expression
%avoids explicitly the problem with the combinatorial explosion in the number of 
%terms of Wick's theorem \cite{PhysRevA.72.032720} which is hidden in the
%determinant of the product of matrices. 
%
A similar expression has been obtained for the more general kind of product wave functions of the Hartree-Fock-Bogoliubov
(HFB) type \cite{Ber12} using pfaffians \cite{Robledo2009}. 

In the right hand side of 
Eq. \eqref{eq:gwt} a potential source of problems is identified in the
inverse of $S_\ab$. If the inverse exists, then $\det S_\ab \neq 0$ and 
it is possible to write
\begin{equation}
  \frac{
  \bra<A| f_M \cdots f_1 g^\dagger_1 \cdots g^\dagger_M \ket|B>}
  {\<A|B\>} = \det\(   S_\fg-S_\fb S_\ab^{-1} S_\ag\)
  \label{eq:gwt1}
\end{equation}
which is the canonical form of the GLT where the sum of $N!$
contractions is replaced by the evaluation of the determinant of a
$M\times M$ matrix. On the other hand, the result of
Eq. \eqref{eq:gwt} is required to resolve the implicit indeterminacy
when $\det S_\ab = 0$ and $S_\ab$ is not invertible (see below).

The above derivation assumes that the $f$ and $g^{\dagger}$ are in
normal order. If this is not the case, operators can always be brought
to normal order using commutation relations of fermion operators. To
illustrate the procedure and to obtain a compact expression we
evaluate now the overlap of a one-body operator $\hat{Q}$
\begin{equation}
  \bra<A| f_M \cdots f_1 \hat{Q} g^\dagger_1 \cdots g^\dagger_M \ket|B>
\label{eq:over_Q}
\end{equation}
where $\hat{Q}$ is written in terms of fermion operators
$r^{\dagger}_m$ and $t_n$ as
\begin{equation}
  \hat{Q} = \sum_{m,n} Q_{mn} r^{\dagger}_m t_n.
\label{eq:def_Q}
\end{equation}
The matrix element $ \bra<A| f_M \cdots f_1 r^{\dagger}_m t_n
g^\dagger_1 \cdots g^\dagger_M \ket|B>$ is evaluated by using the
commutation relation $r^{\dagger}_m t_n= - t_n r^\dagger_m +
\(S_{\text{tr}}\)_{nm}$ as
\begin{align}
\bra<A| f_M \cdots f_1 r^{\dagger}_{m} t_{n} g^\dagger_1 \cdots g^\dagger_M \ket|B> = \nonumber \\
  \(S_{\text{tr}}\)_{mn}\det\mat S_\fg,S_\fb,S_\ag,S_\ab. - 
  \det\matt S_{\text{tr}},S_{\text{tg}},S_{\text{tb}},S_{\text{fr}},S_\fg,S_\fb,S_{\text{ar}},S_\ag,S_\ab.
\label{eq:der1}
\end{align}
With obvious notation, we introduce the row  $S_{\text{t,gb}}$ and column
$S_{\text{fa,r}}$ vectors as well as the matrix $S_{\text{fa,gb}}$ to be able to use
property \eqref{eq:det_block}. Straightforward 
manipulations lead to the final expression
\begin{align}
  \bra<A| f_M \cdots f_1 r^{\dagger}_{m} t_{n} g^\dagger_1 \cdots g^\dagger_M \ket|B> = \nonumber \\
  S_{\text{t,gb}} S_{\text{fa,gb}}^{-1} S_{\text{fa,r}} \det\( S_{\text{fa,gb}} \)
\label{eq:def_Qme}
\end{align}
For the evaluation of the overlap of a two body operator the matrix element
\begin{equation}
\bra<A| f_M \cdots f_1 r^{\dagger}_{m_1} r^{\dagger}_{m_2} t_{n_1}t_{n_2} g^\dagger_1 \cdots g^\dagger_M \ket|B>
\label{eq:over2}
\end{equation}
is required. Using the same procedure as before and after a few
manipulations we obtain 
\begin{align}
\bra<A| f_M \cdots f_1 r^{\dagger}_{m_1} r^{\dagger}_{m_2} t_{n_1}t_{n_2} g^\dagger_1 \cdots g^\dagger_M \ket|B> = \nonumber \\
\det \( S_{\text{t,gb}} S_{\text{fa,gb}}^{-1} S_{\text{fa,r}} \)
\det \( S_{\text{fa,gb}} \)
\label{eq:def_Vme}
\end{align}
where $S_{\text{t,gb}}$ and $S_{\text{fa,r}}$ have dimensions $2\times 
(M+N)$  and $(M+N)\times 2$, respectively. The matrix element is given 
by the product of $ \det ( S_{\text{fa,gb}} ) $ times the determinant 
of a $2\times 2$ matrix with entries corresponding to the ``elementary 
contractions". The generalization to more general $k$-particle, 
$k$-hole matrix elements is straightforward and leads to the determinant 
of a $k\times k$ matrix of contractions. The 
combinatorial increase in the number of terms as $k$ increases is thus 
hidden in the form of a determinant of low dimensionality. This result 
is the generalization of Eq.~(51) of \cite{PhysRev.97.1474}.

To finish this section let us consider a common situation concerning
to symmetry restoration where the overlap includes a multi-particle
unitary operator $\hat{\mathcal{T}}$ in the form of an exponentiated one
body-operator. Typical examples are the rotation and translation operator. Then the
operator $b^\dagger_m$ generating the $|B\rangle$ configuration are transformed to
$\tilde b^\dagger_m$ given by

\begin{equation}
  \hat{\mathcal{T}} b^{\dagger}_m \hat{\mathcal{T}}^\dagger =
  \tilde{b}^{\dagger}_m = \sum_n \( T_b \)_{nm} b^\dagger_n
\label{eq:TbT}
\end{equation}
The overlap becomes
\begin{equation}
\bra<A| f_M \cdots f_1  g^\dagger_1 \cdots g^\dagger_M \hat{\mathcal{T}} \ket|B> =
  \det\mat S_\fg,S_{\text{f\~b}},S_\ag,S_{\text{a\~b}}.
\label{eq:over_T}
\end{equation}
where the only modification with respect to Eq. \eqref{eq:gwt_det} is in the 
overlaps $S_{\text{f\~b}}$ and $S_{\text{a\~b}}$ which have to be computed
with the $\tilde{b}^{\dagger}_m$ of Eq. \eqref{eq:TbT}.

% ---------------------------------------------------------------------------
%                                               Z E R O    O V E R L A P 
% ---------------------------------------------------------------------------

\section{Case of zero overlap}
\label{sec:zero}

Let us study how to apply the GLT of Eq.~(\ref{eq:gwt}) when the overlap
between the states $\ket|A>$ and $\ket|B>$ is zero. The methodology
used can be used straightforwardly for the other form of Löwdin's
theorem, Eq.~(\ref{eq:def_Vme}). When $\<A|B\>=0$, $S_{ab}$ is a
singular matrix and Eq. \eqref{eq:gwt} becomes indeterminate. To avoid
the problem one can always use the full determinant in Eq.
\eqref{eq:gwt_det}, of order $(N+M)\times(N+M)$, but this comes at a
higher cost than just using \eqref{eq:gwt}. In addition, resolving the
indeterminacy explicitly is always beneficial in order to avoid
numerical artifacts that could eventually appear. To this end, we introduce
the {\em singular value decomposition} (SVD) of $S$
\begin{equation}
  S_\ab=U \Sigma V^\dagger,
  \label{eq:svd}
\end{equation}
where $U$ and $V$ are orthogonal matrices ($U^\dagger U=I$) and
$\Sigma$ is diagonal. If $S_\ab$ is near singular, it can be expressed as
\def\eps{\varepsilon}
\begin{equation}
  \Sigma = \begin{pmatrix}
    \sigma_1 & \\
        & \ddots & \\
        &        & \sigma_{N-k} \\
        &        &         & \eps_1 \\
        &        &         &       & \ddots & \\
        &        &         &       &        & \eps_k
  \end{pmatrix} \equiv \begin{pmatrix} \Sigma^\textrm{R} & \\ & E \end{pmatrix}
\label{eq:sigma}
\end{equation}
with $\eps_i$ a set of $k$ small numbers, 
while $E$ is the $k\times k$ diagonal matrix with $\eps_i$ in the
diagonal. Using this decomposition we arrive at
\begin{equation}
  \det S_\ab= f \det \Sigma,
\end{equation}
where  $f=e^{i\varphi_{UV}}\equiv \det U \det V^\dagger$.
%and
%
%\begin{equation}
%  \det\Sigma=\det E \prod_{i=1}^{N-k} \sigma_i=\det E \det\Sigma^\textrm{reg},
%\end{equation}
%
One also has
\begin{equation}
  S_\ab^{-1}=V \Sigma^{-1} U^\dagger.
\end{equation}
Let us define now $S^V_\fb\equiv S_\fb V$ and $S^U_\ag\equiv
U^\dagger S_\ag$, and decompose them in a {\em regular} (R) and a
{\em singular} (S) part, according to the decomposition in
Eq. \eqref{eq:sigma}

\begin{equation}
  S^V_\fb= \left.\begin{pmatrix}
    \undermat{N-k}{\bar S^{V,\textrm{R}}_\fb} &
    \undermat{k}{\bar  S^{V,\textrm{S}}_\fb}
  \end{pmatrix} \right\rbrace \textrm{\scriptsize\em M}
\end{equation}
and 
\begin{equation}
  S^U_\ag = \left(
  \begin{array}{c}
    \bar S^{U,\textrm{R}}_\ag \\[2mm]
    \undermat{M}{\bar S^{U,\textrm{S}}_\ag} \\
    \end{array}
  \right)
    \begin{array}{l}
    \left.\right\rbrace \textrm{\scriptsize\em N-k} \\[2mm]
    \left.\right\rbrace \textrm{\scriptsize\em k} \\
    \end{array}
\end{equation}
\vspace{3mm}

Using this decomposition we can write
\begin{equation}
  \det\mat S_\fg,S_\fb,S_\ag,S_\ab. =
  f \det \Sigma \det \( S_\fg - S^V_\fb \Sigma^{-1} S^U_\ag  \),
\end{equation}
with
\begin{align}
  S^V_\fb \Sigma^{-1} S^U_\ag = &
   \bar S^{V,\textrm{R}}_\fb \(\Sigma^{\textrm{R}}\)^{-1} \bar S^{U,\textrm{R}}_\ag 
   \nonumber \\ 
   +& \bar S^{V,\textrm{S}}_\fb E^{-1} \bar S^{U,\textrm{S}}_\ag,
  \end{align}
Let us now introduce the $M\times M$ matrix
\begin{equation}
  C \equiv S_\fg - \bar S^{V,\textrm{R}}_\fb \( \Sigma^{\textrm{R}} \)^{-1} \bar
  S^{U,\textrm{R}}_\ag,
\end{equation}
and consider the determinant
\begin{equation}
  \det \( C-\bar S^{V,\textrm{S}}_\fb E^{-1} \bar S^{U,\textrm{S}}_\ag \).
\end{equation}
It can be computed using the Woodbury formula for the determinant 
(see, for instance, Ref \cite{harville2008matrix})
\begin{equation}
  \det\( A+UWV^\dagger \) = \det\(W^{-1}+V^\dagger A^{-1} U\) \det W
  \det A
\end{equation}
to obtain
\begin{align}
  &\det\mat S_\fg,S_\fb,S_\ag,S_\ab. =
  f \det\Sigma^\textrm{R} \det E \times\nonumber\\
 & \times
  \det\( E - \bar S^{U,\textrm{S}}_\ag C^{-1}\bar S^{V,\textrm{S}}_\fb \)
  \frac{1}{\det E} \det C,
\end{align}
which is well defined in the limit $\eps_i\to 0$
\begin{equation}
  \det\mat S_\fg,S_\fb,S_\ag,S_\ab. = f\det\Sigma^\textrm{R}
  \det\(  -\bar S^{U,\textrm{S}}_\ag C^{-1}\bar S^{V,\textrm{S}}_\fb \) \det C,
  \label{eq:gwt_zover}
\end{equation}
which is a finite quantity when $\eps_i\to 0$. This quantity requires the 
SVD of $S_\ab$ to get $\Sigma^\textrm{R}$ and the $U$ and $V$ matrices.
For large values of $N$ this can be a costly operation of order $N^{3}$.
The inverse of the diagonal $\Sigma^\textrm{R}$ is also required
as well as the construction of the  $S^{U}_\ag$ and
$S^{V}_\fb$ matrices. Once we have all the ingredients, the evaluation
of the determinants in Eq. \eqref{eq:gwt_zover} require little extra work
due to the low dimensionality of the matrices involved ($k \times k$ and
$M\times M$). Please note that the dimensionality of the different matrices
appearing in Eq. \eqref{eq:gwt_zover} is not the same ($\Sigma^\textrm{R}$
is $(N-k)\times(N-k)$, $\bar S^{U,\textrm{S}}_\ag C^{-1}\bar S^{V,\textrm{S}}_\fb$
is $k\times k$ and $C$ is $M\times M$) and therefore the formula for the product
of a determinant does not apply here. 
    
Note that the derivation above can also be extended to the case where
the matrix $S_\ab$ is ill-conditioned and has a very small (but
non-zero) determinant. A blind use of the traditional
formulas may contaminate the final results due to numerical artifacts
consequence of the finite representation of floating point numbers in
computer's arithmetic.

%%%%%%%%%%%%%%%%%%%%%%%%%%%%%%%%%%%%%%%%%%%%%%%%%%%%%%%%%%%%%%%%%%%%%%%%%%%

\section{Numerical experiments}
\label{sec:numerical}

In this section we put the extended Löwdin approach to the test,
characterizing the entanglement behavior of a linear combination of
two Slater determinants.

\subsection{The rainbow system}

\begin{figure}
  \includegraphics[width=8cm]{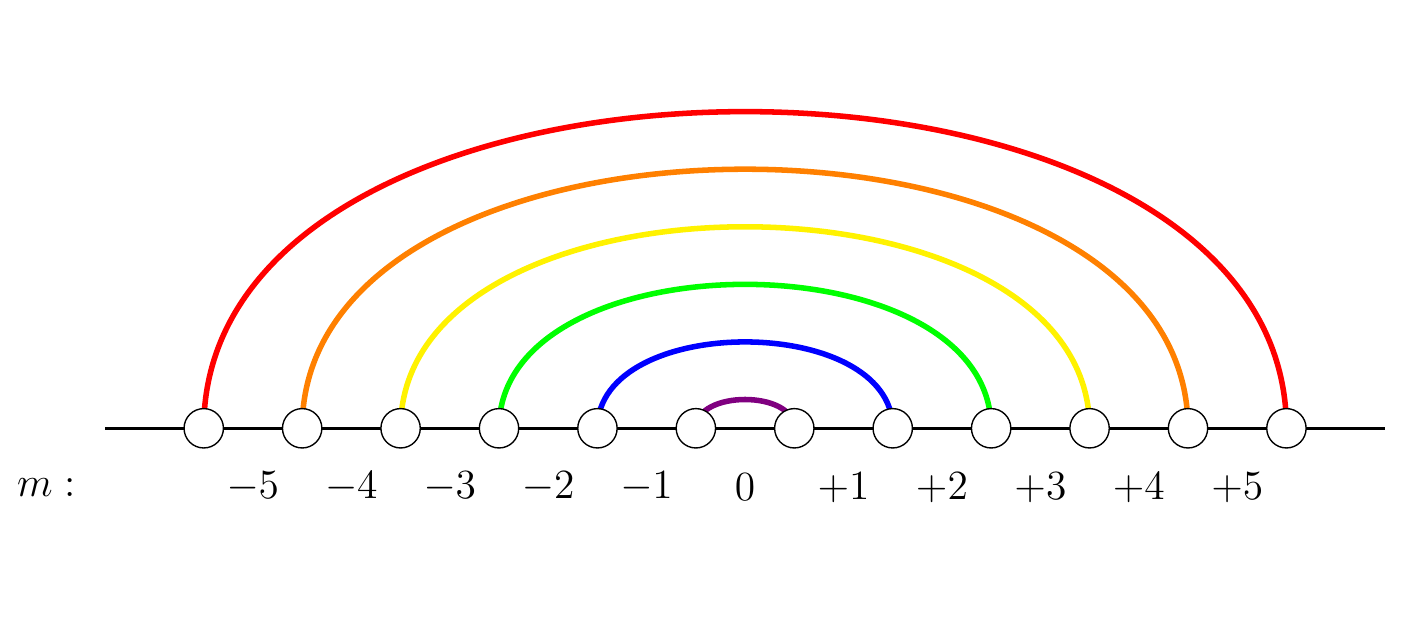}
  \caption{Illustration of the rainbow ground state in the small
    $\alpha$ limit: a valence bond solid with concentric bonds around
    the center of the chain. Notice that the entanglement entropy of a
    left block is proportional to the block size, i.e. a volumetric
    growth of the entropy.}
  \label{fig:illust}
\end{figure}

As our physical system, we have chosen the {\em rainbow system}
\cite{Vitagliano.10,Ramirez.14,Ramirez.15}, a 1D inhomogeneous
fermionic hopping system which presents volumetric entanglement
between its left and right halves. It can be described on an open
chain through the following Hamiltonian,

\beq
H=-\sum_{m=-L+1}^{L-1} J_m c^\dagger_m c_{m+1} + \text{h.c.},
\label{eq:hamrainbow}
\eeq
with hopping amplitudes given by

\beq
J_m = \begin{cases}
  \alpha^{2m+1} & \text{if $m\neq 0$,} \\
  1& \text{otherwise,}\\
\end{cases}
\label{eq:hoprainbow}
\eeq
in terms of an inhomogeneity parameter $\alpha \in (0,1]$.
For $\alpha=1$, the system reduces to the homogeneous case. For small
$\alpha$, the ground state (GS) of Hamiltonian \eqref{eq:hamrainbow}
becomes approximately a {\em valence bond solid} with concentric bonds
around the center, see Fig. \ref{fig:illust}. This GS violates
maximally the area-law \cite{Eisert.10}, giving rise to a volumetric
growth of entanglement.

\subsection{Entanglement and number fluctuations}

The eigenstates of Hamiltonian \eqref{eq:hamrainbow} are Slater
determinants. Therefore, the exact entanglement entropy for any block
can be efficiently evaluated \cite{Peschel.03}. Yet, this technique of
evaluating the entanglement entropy cannot be extended to a linear
combination of Slater determinants.

Let us consider an arbitrary linear combination of the ground state
and first excited states of Hamiltonian \eqref{eq:hamrainbow}, both
within the half-filling sector, i.e. with $N/2$ particles.

\beq
\ket|\Psi>=\alpha\ket|0>+\beta\ket|1>.
\label{eq:twoslater}
\eeq
We are interested in estimates of the entanglement between a region
$A$ and its complementary, always within state $\ket|\Psi>$. The
procedure is straightforward in the case of a single Slater
determinants \cite{Peschel.03}. In that case, the computation of the
full spectrum of the reduced density matrix can be performed in
$O(N^k)$ operations, with a low value of $k$. On the other hand, the
evaluation of the entanglement properties for a generic state takes,
in principle, $O(2^N)$ operations. Yet, there are some observables
which act as {\em entanglement witnesses}, i.e.: whose values act as
an indicator of entanglement. The fluctuation of the number of
particles is one of these observables \cite{Klich.06}. 

Let $n_A$ be the number of particles on region $A$:

\beq
n_A=\sum_{i\in A} n_i = \sum_{i\in A} c^\dagger_i c_i,
\label{eq:nA}
\eeq
whose fluctuations are given by

\beq
\sigma_N^2 = \< n_A^2 \> - \<n_A\>^2.
\label{eq:dev}
\eeq
The expectation value of the number of particles on region $A$ is
found through the following calculation:

\begin{align}
  \<n_A\> &=\sum_{i\in A} \elem<\Psi|n_i|\Psi> \nonumber \\
  =&
\sum_{i\in A} \[ |\alpha|^2 \elem<0|n_i|0> + |\beta|^2 \elem<1|n_i|1> \right.
\nonumber \\
 +& \left. 2\text{Re}\;\bar\alpha\beta \elem<0|n_i|1> \].
\label{eq:na}
\end{align}
The first two terms are easily found, because they refer to a single
Slater determinant. The third, notwithstanding, must be found using
the (generalized) Löwdin tricks described:

\beq
\elem<0|n_i|1>=\elem<0|c^\dagger_i c_i|1>=\<0|1\> -
\elem<0|c_ic^\dagger_i|1>.
\label{eq:0n1}
\eeq

The quadratic term is more involved:

\begin{align}
\<n_A^2\>= &\sum_{i,j\in A} \[|\alpha|^2\elem<0|n_in_j|0>
+|\beta|^2\elem<1|n_in_j|1> \right. \nonumber \\
& \left. + 2 \text{Re}\; \bar\alpha \beta \elem<0|n_in_j|1> \].
\label{eq:na2}
\end{align}

Again, the first two terms are straightforward to obtain. If $C_{A,0}$ is
the submatrix of the correlation matrix corresponding to block $A$ on
state $\ket|0>$, then

\beq
\elem<0|n_in_j|0>=\Tr(C_{A,0})^2+\Tr(C_{A,0}(I-C_{A,0})).
\label{eq:0ninj0}
\eeq

The last term of Eq. \eqref{eq:na2} is the most involved one, because
we cannot assume Wick's theorem. We find

\begin{align}
\elem<0|n_in_j|1>&= \elem<0|(1-c_ic^\dagger_i)(1-c_jc^\dagger_j)|1>= \nonumber\\
&= \<0|1\> - \elem<0|c_jc^\dagger_j|1> - \elem<0|c_ic^\dagger_i|1> \nonumber\\
&+ \delta_{ij} \elem<0|c_ic^\dagger_j|1>
-\elem<0|c_ic_jc^\dagger_ic^\dagger_j|1>.
\end{align}

It has been proved that, for a single Slater determinant
\cite{Klich.06},

\beq
S_A\geq 4\log 2\sigma_N^2.
\label{eq:klich}
\eeq
Thus, even though Eq. \eqref{eq:klich} is not proved for a generic
linear combination of Slater determinants, we will employ $\sigma_N$
as our estimate for an {\em entanglement witness}.

\subsection{Numerical experiments}

For concreteness, let us set $\alpha=\sqrt{x}$ and $\beta=\sqrt{1-x}$ for
$x\in [0,1]$ in Eq. \eqref{eq:twoslater}. Thus, our state will be
given by 

\beq
\ket|\Psi> = \sqrt{x} \ket|0> + \sqrt{1-x} \ket|1>,
\label{eq:state}
\eeq
Let us notice that the first excitation is obtained from the ground
state by performing a parity transformation on the Fermi level
\cite{Ramirez.15}. Notice that, by construction $\<0|1\>=0$, thus
forcing us to make all our computations in the {\em zero overlap} case.

\begin{figure}
  \includegraphics[width=8cm]{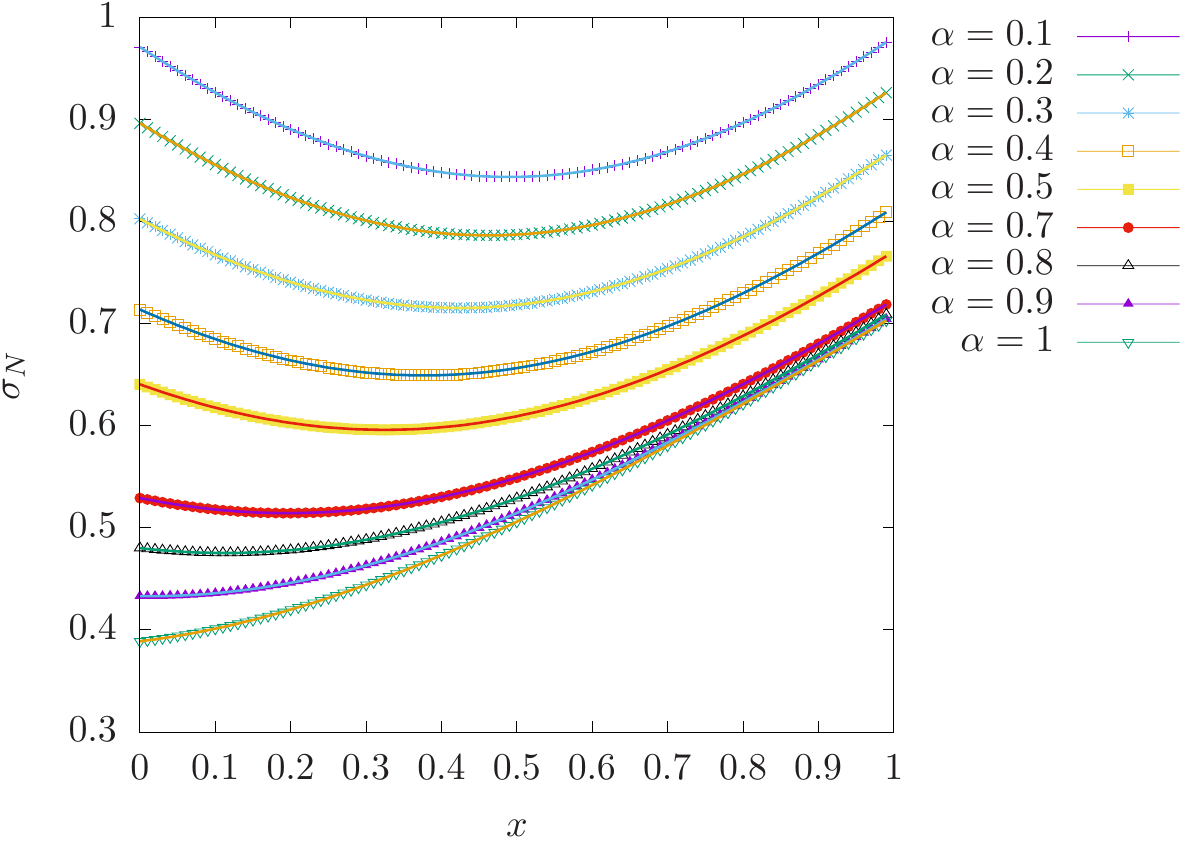}
  \includegraphics[width=8cm]{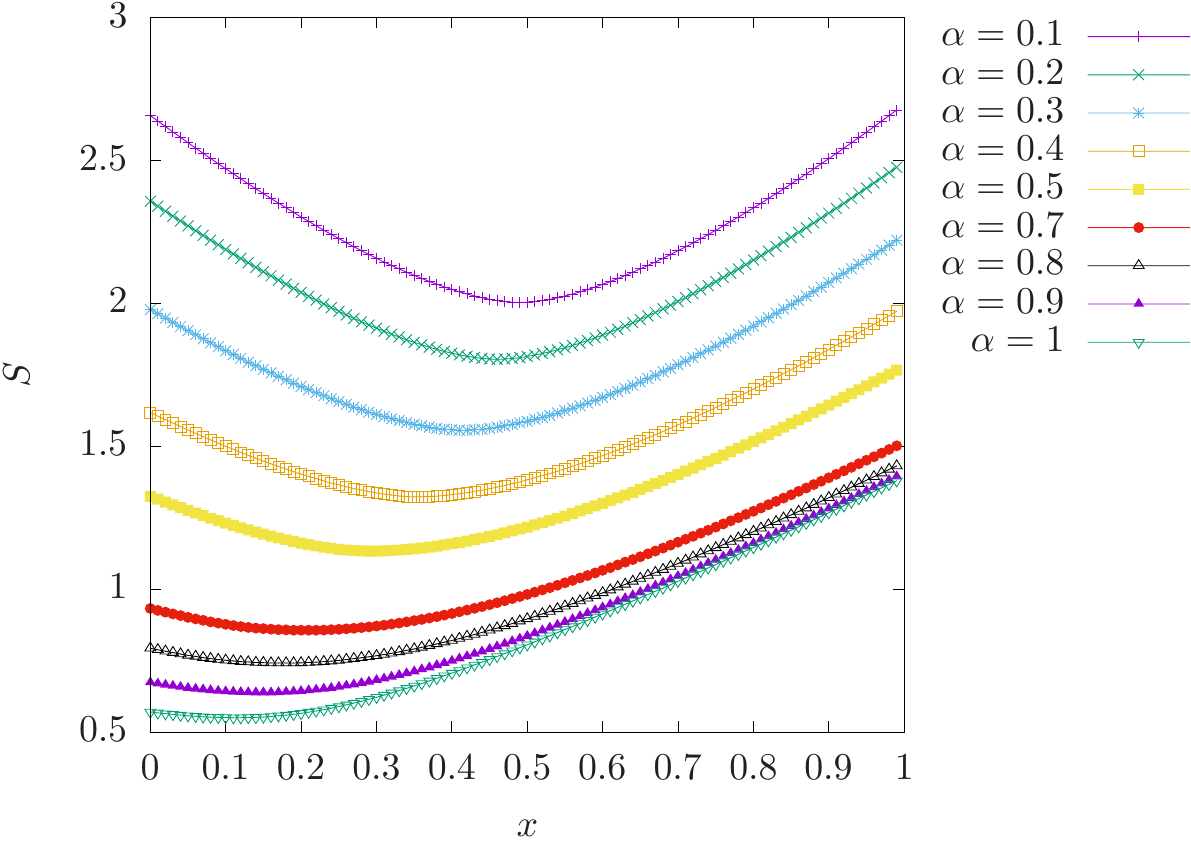}
  \caption{Top: deviation of the number of particles in the left half
    of state \eqref{eq:state} with $N=8$ sites and various values of
    $\alpha$, computed with the full Slater determinants and with our
    generalized Löwdin scheme. The theoretical value in the $\alpha\to
    0^+$ limit is $\sigma_A=1$. Bottom: entanglement of the left half
    of state \eqref{eq:state}, computed with the full Slater
    determinants. The theoretical value in the $\alpha\to 0^+$ limit
    is $4\log 2\approx 2.77$.}
  \label{fig:N8}
\end{figure}

In the top panel of Fig. \ref{fig:N8} we show the variance $\sigma_N$ of the
number of particles in the left half of the rainbow system with $N=8$,
for different values of $\alpha$. Notice that only in the
$\alpha\to 0^+$ limit the variance is the same for $x=0$ and $x=1$,
i.e.: in the GS and the first excited. This variance has been computed
in two different ways: the dots correspond to the exact calculation,
with the full Slater determinant, and the continuous line corresponds
to the computation performed with the generalized Löwdin formulas derived
in this article. The agreement is complete, and the
computational time is enormously reduced with our tools.

In the $\alpha\to 0^+$ limit, the left-half of the rainbow system
becomes an infinite temperature mixed state. Thus, the fluctuations in
the particle number are easy to obtain, following a binomial
distribution, $\sigma_A = \sqrt{N/8}$, that we can readily check in
Fig. \ref{fig:N8}. Also, the entanglement entropy will grow up to the
maximal possible value, $(N\log 2)/2$.

\begin{figure}
  \includegraphics[width=8cm]{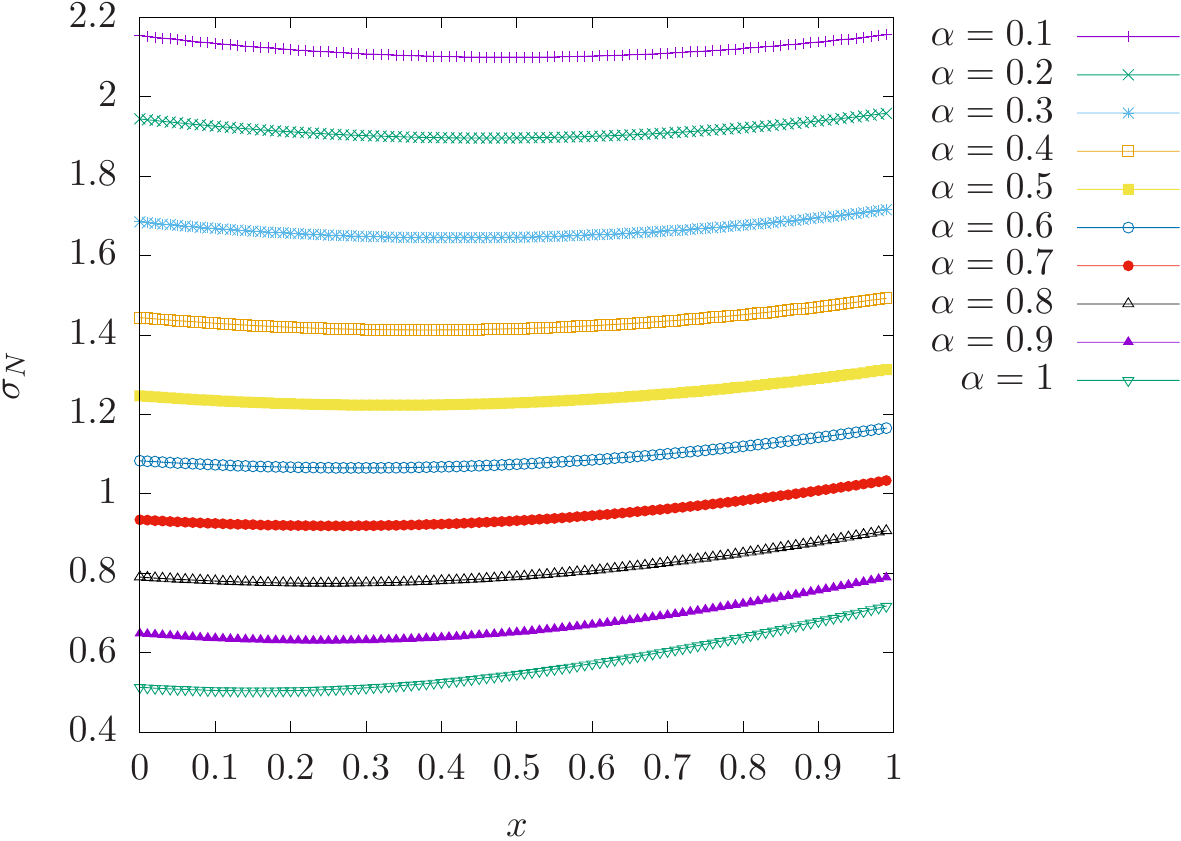}
  \caption{Same observable as Fig. \ref{fig:N8} for a rainbow system
    with $N=40$. The theoretical value for the deviation of the
    particle number of the left half in the $\alpha\to 0^+$ limit is
    $\sqrt{N/8}=\sqrt{5}\approx 2.23$.}
  \label{fig:N40}
\end{figure}

The lower panel of Fig. \ref{fig:N8} shows the Von Neumann entropy for
the same blocks as the top panel. Notice that, in this case, our
calculations cannot be easily extended. Our generalized Löwdin
calculation scheme is much more efficient, and can be extended to
larger system sizes. Fig. \ref{fig:N40} shows the deviation in the
number of particles of state \eqref{eq:state} as a function of $x$ for
various values of $\alpha$ in a chain with $N=40$ sites.

\subsection{Generalized Löwdin C++ code}

We have uploaded our C++ libraries to compute matrix elements of
Slater determinants using the generalized Löwdin approach to the
repository {\tt github} as {\em free software} \cite{github}. The same
material can also be downloaded from the Supplementary material section of
the electronic version of the journal \cite{SuppM}.

%%%%%%%%%%%%%%%%%%%%%%%%%%%%%%%%%%%%%%%%%%%%%%%%%%%%%%%%%%%%%%%%%%%%%%%%%%%%

\section{Conclusions and further work}

We have extended the seminal results of Löwdin to obtain efficiently
the matrix elements of an arbitrarily large product of fermionic
operators between arbitrary Slater determinants. Our results are still
applicable when the overlap matrix between the orbitals of the Slater
determinants is singular, i.e. when the corresponding states are
orthogonal.

Efficient computation of matrix elements in non-orthogonal Slater
determinants will open a very interesting possibility: the creation of
Ansätze including Slater determinants obtained from different
procedures and, therefore, using different orbitals.

As proposals of future work, we would like to remark the extension of
the previous calculations to the obtention of the full reduced density
matrix, combining our results with those of \cite{Peschel.03} for the
reduced density matrix of a block of a single Slater determinant.

%%%%%%%%%%%%%%%%%%%%%%%%%%%%%%%%%%%%%%%%%%%%%%%%%%%%%%%%%%%%%%%%%%%%%%%%%%%%

\acknowledgments{}
We thank G.F. Bertsch for a careful reading of the manuscript and 
suggestions. This work has been supported by the Spanish Ministerio de 
Ciencia, Innovaci\'on y Universidades and the European regional 
development fund (FEDER), grants Nos FIS2015-69167-C2-1 (JRL), 
FIS2015-63770 (JD and LMR), PGC2018-094180-B-I00 (JD), and  
FPA2015-65929, PGC2018-094583-B-I00 (LMR).

\bibliographystyle{apsrev4-1}
%\bibliography{overlapsSlater}
%

\end{thebibliography}
\end{document}